\documentclass[twocolumn,showpacs,preprintnumbers,amssymb]{revtex4-1}
\usepackage{graphicx}
\usepackage{dcolumn}
\usepackage{color}
\usepackage{bm}
\usepackage{amsmath}
\usepackage{amssymb}
\usepackage{epsfig}

\def \a{\alpha}
\def \b{\beta}
\def \l{\lambda}
\def \L{\Lambda}
\def \e{\epsilon}
\def \g{\gamma}
\def \d{\delta}
\def \k{\kappa}

\def \be{\begin{equation}}
\def \ee{\end{equation}}
\def \ben{\begin{eqnarray}}
\def \een{\end{eqnarray}}

\def \O{\Omega}

\def \S{\Sigma}
\def \p{\partial}
\def \t{\theta}
\def \P{\Phi}
\def \M{\mathcal{M}}
\def \G{\mathcal{G}}
\def \r{\rho}
\def \k{\kappa}

\begin{document}

\title{Gravitational Collapse for the {\bf K-}essence Emergent Vaidya Spacetime}

\author{Goutam Manna }
\altaffiliation{goutammanna.pkc@gmail.com}
\affiliation{Department of Physics, Prabhat Kumar College, Contai, Purba Medinipur-721404, India}

\begin{abstract}

In this paper, we study the gravitational collapse in the {\bf k-}essence emergent gravity using a generalized Vaidya-type metric as a background. We also analyze the cosmic censorship hypothesis for this system. We show that the emergent gravity metric resembles closely to the new type of the generalized Vaidya metrics for null fluid collapse with the {\bf k-}essence emergent mass function, where we consider the
{\bf k-}essence scalar field being a function solely of the advanced or the retarded time. This new type of {\bf k-}essence
emergent Vaidya metric has satisfied the required energy conditions. The existence of the locally naked central singularity, the strength and the strongness of the singularities for the {\bf k-}essence emergent Vaidya metric are the interesting outcomes of the present work.

\pacs{04.20.-q, 04.20.Dw, 04.70.Bw }

\keywords{K-essence, Emergent Gravity, Gravitational Collapse, Vaidya Spacetime, Cosmic censorship}

\end{abstract}

\maketitle

\section{Introduction}
\label{intro}
The theoretical evolution of the gravitational collapse \cite{joshi1,joshi2} guided by
the general theory of relativity leads to either a black hole or a
naked singularity, which can communicate with far-away observers in the universe. The radiating Schwarzschild
spacetime, also known as the Vaidya spacetime \cite{vaidya1}, describes the geometry outside a radiating spherically symmetric star.
In particular, Papapetrou \cite{papa} first showed that the solution of null dust fluid with spherical symmetry in gravitational collapse can give rise to the formation of
naked singularities. It provides one of the  counter examples to the cosmic censorship conjecture (CCC) \cite{penrose}. In \cite{joshi3,vert}, they have described the causal trajectories joining the singularities in the ingoing Vaidya situation. Also a relatively complete description
of constraints is provided classifying the non-spacelike
geodesics that connect the naked singularity in the past. It is then shown to be a strong curvature singularity
in a more substantial sense. The generalization of the Vaidya solution \cite{vai1,vai2,vai3,vai4,vai5,jg}, includes all the known solutions of Einstein’s field equations with a combination of Type-I and Type-II matter fields. It was given
by Husain \cite{husain} and
Wang and Wu \cite{wang}. The generalization of the Vaidya solution is
also known as the generalized Vaidya spacetime.
In the context of the cosmic censorship
hypothesis, the gravitational collapse for the generalised
Vaidya spacetime has been studied in \cite{maombi}. They showed that
the classes of generalized Vaidya mass functions have come in the scenario
indicating the termination of collapse with a locally naked
central singularity. They calculated the strength of these
singularities.  A general mathematical framework was developed to study the conditions on the mass function so that non-spacelike geodesics directed towards future can terminate at the singularity in the past. They also exhibited that, for a given generalized Vaidya mass function,
how the final fate of collapse is determined in terms of either a black hole or a naked singularity. In \cite{patil}, the gravitational collapse of higher-dimensional is discussed in the charged-Vaidya spacetime. They have shown that singularities arising in a charged null fluid in higher dimension.
These singularities are
always naked, violating strong cosmic censorship hypothesis (CCH).  This hypothesis is not necessarily on weak cosmic censorship. In \cite{cag,momeni}, the holographic complexity and the holographic entropy cone in AdS-Vaidya spacetime have been explored. The attempt to revisit Vaidya-Tikekar
stellar model in the linear regime has
been explored in \cite{sharma}.

To be more specific on the {\bf k-}essence model, we note that it is a scalar field model. In this model, the kinetic energy of the field dominates over the potential energy of the field. The theoretical form of the {\bf k-}essence field Lagrangian is non-canonical. The Lagrangian cannot be separated
in the terms corresponding to kinetic energy and potential energy. Also,
it does not depend explicitly on the field itself. A theory with a non-canonical kinetic term was first proposed by
Born and Infeld \cite{born1,born2,born3}. It was proposed to get rid of the problem related to the infinite self-energy of the electron. The general form of the lagrangian for {\bf k-}essence model is: $L=-V(\phi)F(X)$ where $\phi$ is {\bf k-}essence scalar field, $X=\frac{1}{2}g^{\mu\nu}\nabla_{\mu}\phi\nabla_{\nu}\phi$ and does not depend explicitly on $\phi$ to start with \cite{babi1,babi2,babi3,babi4,babi5,gm1,gm2,gm3,scherrer1,scherrer2}. The Lagrangian for {\bf k-}essence scalar fields contain non-canonical kinetic terms.
The differences between the {\bf k-}essence theory with non-canonical kinetic
terms and the relativistic field theories with canonical kinetic terms lie in the non-trivial dynamical solutions
of the {\bf k-}essence equation of motion (EOM). The solutions of the EOM not only spontaneously breaks the Lorentz invariance but also changes the
metric for the perturbations around these solutions. Thus the perturbations propagate in the so-called {\it emergent or analogue} curved spacetime \cite{babi1,babi2,babi3,babi4,babi5} with the metric which is different from the gravitational one.  The emergent gravity metric $\tilde G_{\mu\nu}$ is not conformally equivalent to the gravitational metric $g_{\mu\nu}$. 


In \cite{ali1}, the gravitational collapse
is analyzed in the context of gravity's rainbow \cite{magueijo}. The rainbow gravity is basically a possible step towards repairing the rifts between the theories of general relativity and quantum mechanics. In \cite{heyd1,heyd2}, they have studied the energy dependent deformation and time dependent geometry of massive gravity using the formalism of massive gravity's rainbow. Both the Vainshtein and the dRGT mechanisms are used for the
energy dependent massive gravity (Vaidya spacetime). Relevant theories of rainbow gravity and their subsequent explorations in Charged dilatonic black holes, Gauss-Bonnet gravity, Lovelock gravity, combination of Rastall and rainbow theories, $AdS_{4}$ dyonic black holes, Deformed Starobinsky model, Thermodynamics of black holes, Galileon gravity, Horizon effect, Violation of weak cosmic censorship and Branes are addressed in \cite{ali2,ali3,ali4,hendi1,hendi2,hendi3,hendi4,hendi5,hendi6,mota,pana,rudra,ashour,chan,gim,haldar}. Also, we clearly mention that in this work we cannot study the Rainbow gravity in the framework of the {\bf k-}essence emergent Vaidya spacetime.

In this paper, we explore the gravitational collapse for the {\bf k-}essence emergent Vaidya spacetime in the context of cosmic censorship hypothesis. Here, we consider the background gravitational metric as the generalised Vaidya metric  \cite{husain,wang,maombi}. 

The paper is organized as follows: The brief review of the {\bf k-}essence theory based on the Dirac-Born-Infeld action is presented in section 2. It makes us familiar
with the so-called emergent gravity spacetime where the gravitational metric is not conformally equivalent to the emergent gravity metric. In the next section, we
construct the {\bf k-}essence emergent Vaidya spacetime for the generalized Vaidya spacetime background. It is accompanied by a scalar field with a restriction of being an arbitrary function of the advanced or retarded Eddington-Finkelstein time. This scalar field is independent of the other variables of the four-dimensional spacetime. We also construct the Ricci and Einstein tensors corresponding to our {\bf k-}essence emergent Vaidya metric. We also compute the components of the emergent energy-momentum tensor by direct substitution into an emergent Einstein equation.
This emergent tensor must obey the energy conditions of the emergent geometry. In section 4, we develope the
collapsing scenario for the {\bf k-}essence emergent Vaidya spacetime. In this section, we analyze the structure of the central singularity to find out the conditions on the
{\bf k-}essence emergent Vaidya mass function, existence of outgoing nonspacelike geodesic and the strength of the
singularities for the above emergent geometry. Some examples of the subclasses of generalized Vaidya backgrounds have also been discussed in section 5. In the
last section, we conclude over the whole work.

\section{Brief Review of K-essence theory and Emergent gravity}
In this section, we narrate a brief description of the development of  construction of the effective metric for the emergent spacetime corresponding to a general background geometry and a very general {\bf k}-essence scalar field sector. The {\bf k}-essence scalar field $\phi$ minimally coupled to the background spacetime metric $g_{\mu\nu}$ has action \cite{babi1}-\cite{babi5}
\ben
S_{k}[\phi,g_{\mu\nu}]= \int d^{4}x {\sqrt -g} L(X,\phi)
\label{eq:1}
\een
where $X={1\over 2}g^{\mu\nu}\nabla_{\mu}\phi\nabla_{\nu}\phi$.
The energy-momentum tensor is
\ben
T_{\mu\nu}\equiv {2\over \sqrt {-g}}{\delta S_{k}\over \delta g^{\mu\nu}}= L_{X}\nabla_{\mu}\phi\nabla_{\nu}\phi - g_{\mu\nu}L
\label{eq:2}
\een
$L_{\mathrm X}= {dL\over dX},~~ L_{\mathrm XX}= {d^{2}L\over dX^{2}},
~~L_{\mathrm\phi}={dL\over d\phi}$ and  
$\nabla_{\mu}$ is the covariant derivative defined with respect to the gravitational metric $g_{\mu\nu}$.
The scalar field equation of motion is
\ben
-{1\over \sqrt {-g}}{\delta S_{k}\over \delta \phi}= \tilde G^{\mu\nu}\nabla_{\mu}\nabla_{\nu}\phi +2XL_{X\phi}-L_{\phi}=0
\label{eq:3}
\een
where  
\ben
\tilde G^{\mu\nu}\equiv L_{X} g^{\mu\nu} + L_{XX} \nabla ^{\mu}\phi\nabla^{\nu}\phi
\label{eq:4}
\een
and $1+ {2X  L_{XX}\over L_{X}} > 0$.

 We adopt the conformal transformation
$G^{\mu\nu}\equiv {c_{s}\over L_{x}^{2}}\tilde G^{\mu\nu}$, with
$c_s^{2}(X,\phi)\equiv{(1+2X{L_{XX}\over L_{X}})^{-1}}$. Then the inverse metric of $G^{\mu\nu}$ takes shape as 
\ben G_{\mu\nu}={L_{X}\over c_{s}}[g_{\mu\nu}-{c_{s}^{2}}{L_{XX}\over L_{X}}\nabla_{\mu}\phi\nabla_{\nu}\phi] .
\label{eq:5}
\een
A further conformal transformation \cite{gm1,gm2} $\bar G_{\mu\nu}\equiv {c_{s}\over L_{X}}G_{\mu\nu}$ gives
\ben \bar G_{\mu\nu}
={g_{\mu\nu}-{{L_{XX}}\over {L_{X}+2XL_{XX}}}\nabla_{\mu}\phi\nabla_{\nu}\phi}
\label{eq:6}
\een	
Here one must always have $L_{X}\neq 0$ for $c_{s}^{2}$ to be positive definite and only then equations $(1)-(4)$ will be physically meaningful.

It is clear that, for non-trivial spacetime configurations of $\phi$, the emergent metric $G_{\mu\nu}$ is, in general, not conformally equivalent to $g_{\mu\nu}$. So $\phi$ has the properties different from canonical scalar fields with the local causal structure distinct from those defined with $g_{\mu\nu}$. Further, if $L$ is not an explicit function of $\phi$, the equation of motion $(3)$ reduces to;
\ben
-{1\over \sqrt {-g}}{\delta S_{k}\over \delta \phi}
= \bar G^{\mu\nu}\nabla_{\mu}\nabla_{\nu}\phi=0
\label{eq:7}
\een
We shall take the Lagrangian as $L=L(X)=1-V\sqrt{1-2X}$ with $V$ is a constant. 
This is a particular case of the Dirac-Born-Infeld (DBI) lagrangian  \cite{born1,born2,born3,gm1,gm2,gm3}
\ben
L(X,\phi)= 1-V(\phi)\sqrt{1-2X}
\label{eq:8}
\een
for $V(\phi)=V=constant$~~and~~$kinetic ~ energy ~ of~\phi>>V$ i.e.$(\dot\phi)^{2}>>V$. This is typical for the {\bf k}-essence fields where the kinetic energy dominates over the potential energy. Then $c_{s}^{2}(X,\phi)=1-2X$. For scalar fields $\nabla_{\mu}\phi=\partial_{\mu}\phi$. Thus effective emergent metric (\ref{eq:6}) becomes
\ben
\bar G_{\mu\nu}= g_{\mu\nu} - \partial _{\mu}\phi\partial_{\nu}\phi
\label{eq:9}
\een

Consider the second conformal transformation $\bar G_{\mu\nu}\equiv {c_{s}\over L_{X}}G_{\mu\nu}$.
Following \cite{wald,gm1} the new Christoffel symbols can be related to the old ones by  
\ben
\bar\Gamma ^{\alpha}_{\mu\nu} 
=\Gamma ^{\alpha}_{\mu\nu} + (1-2X)^{-1/2}G^{\alpha\gamma}[G_{\mu\gamma}\partial_{\nu}(1-2X)^{1/2}\nonumber\\
+G_{\nu\gamma}\partial_{\mu}(1-2X)^{1/2}-G_{\mu\nu}\partial_{\gamma}(1-2X)^{1/2}]\nonumber\\
=\Gamma ^{\alpha}_{\mu\nu} -\frac {1}{2(1-2X)}[\delta^{\alpha}_{\mu}\partial_{\nu}X
+ \delta^{\alpha}_{\mu}\partial_{\nu}X]~~~~~~~~~~~
\label{10}
\een
Note that the second term on the right hand side is symmetric under exchange of $\mu$ and $\nu$ 
so that the symmetry of $\bar\Gamma$ is maintained. The second term has its origin solely to the 
{\bf k-}essence lagrangian and this additional term signifies additional interactions (forces). 
The geodesic equation for the {\bf k-}essence theory in terms of the new Christoffel connections  $\bar\Gamma$ now becomes 
\ben
\frac {d^{2}x^{\alpha}}{d\l^{2}} +  \bar\Gamma ^{\alpha}_{\mu\nu}\frac {dx^{\mu}}{d\l}\frac {dx^{\nu}}{d\l}=0
\label{1.11}
\een
where $\l$ is an affine parameter.

\vspace{0.2in}
\section{Emergent Spacetime for the generalized Vaidya Background}
We consider the background gravitational metric as the generalized Vaidya metric. 
The line element for the generalized Vaidya spacetime is \cite{husain,wang,maombi}:
\ben
ds^{2}=-(1-\frac{2m(v,r)}{r})dv^{2}+2dvdr+r^{2}d\O^{2}
\label{12}
\een
where $d\O^{2}=d\t^{2}+d\P^{2}$.
Here $m(v,r)$ is the mass function related to the gravitational energy within a given radius r and $v$ is
the null coordinate corresponding to the Eddington advanced time with $r$ decreasing toward the future along a ray related to $v=constant$.

From (9), the {\bf k-}essence emergent line element can be written as
\ben
dS^{2}=ds^{2}-\p_{\mu}\phi\p_{\nu}\phi dx^{\mu}dx^{\nu}.
\label{eq:13}
\een
We consider the {\bf k-}essence scalar field $\phi(x)=\phi(v)$ only, so that the emergent line element (13) is
\ben
dS^{2}=-\Big[1-\frac{2m(v,r)}{r}-\phi_{v}^{2}\Big]dv^{2}+2 dvdr+r^{2}d\O^{2}~~~
\label{eq:14}
\een
where $\phi_{v}=\frac{\p \phi}{\p v}$ and $\phi_{v}^{2}$ is the kinetic energy of the {\bf k-}essence scalar field which {\it cannot be zero.} Notice that this assumption on $\phi$ actually violates local Lorentz invariance, since in general, spherical symmetry would only require that $\phi(x)=\phi(v,r)$.  The additional assumption of the independence of $\phi$ 
i.e.,  $\phi(v,r)=\phi(v)$  implies that outside of this particular choice of frame, a spherically symmetric $\phi$ is actually a function of both $v$ and $r$. Mention that the {\bf k-}essence theory allows us this type of  Lorentz violation, since the dynamical solutions of the k-essence equation of
motion spontaneously break Lorentz invariance
and also change the metric for the perturbations around these solutions. 

The {\bf k-}essence emergent Vaidya line element (14) also written as
\ben
dS^{2}=-\Big[1-\frac{2\mathcal{M}(v,r)}{r}\Big]dv^{2}+2 dvdr+r^{2}d\O^{2}~~~
\label{eq:15}
\een
where we define the {\bf k-}essence emergent Vaidya mass function 
\ben
\mathcal{M}(v,r)=m(v,r)+\frac{r}{2}\phi_{v}^{2}
\label{16}
\een
which is related to the gravitational energy of the {\bf k-}essence emergent gravity within a given radius $r$.

Now we compute the non-vanishing components of the emergent Ricci tensor and the emergent Einstein's tensor following \cite{wang,maombi}, with the notations having subscripts on $r$ and $v$ denote derivatives with respect to those variables.

Using the following definitions:
\ben
\M_{v}=\dot{\M}(v,r)\equiv \frac{\p\M(v,r)}{\p v}~;~
\M_{r}=\M^{'}(v,r)\equiv\frac{\p\M(v,r)}{\p r}\nonumber\\
\label{17}
\een
the nonvanishing components of the emergent Ricci tensor are
\ben
\bar{R}^{v}_{v}=\bar{R}^{r}_{r}=\frac{\M_{rr}}{r}=\frac{m_{rr}}{r}~;~\nonumber\\
\bar{R}^{\t}_{\t}=\bar{R}^{\P}_{\P}=\frac{2\M_{r}}{r^{2}}=\frac{2m_{r}}{r^{2}}+\frac{\phi_{v}^{2}}{r^{2}}
\label{18}
\een
and the Ricci scalar is given by
\ben
\bar{R}=\frac{2\M_{rr}}{r}
+\frac{4\M_{r}}{r^{2}}
=\frac{2m_{rr}}{r}
+\frac{4m_{r}}{r^{2}}+\frac{2\phi_{v}^{2}}{r^{2}}.
\label{19}
\een
The nonvanishing components of the Einstein's tensor are
\ben
\G^{v}_{v}=\G^{r}_{r}=-\frac{2\M_{r}}{r^{2}}=-\frac{2m_{r}}{r^{2}}-\frac{\phi_{v}^{2}}{r^{2}}~;~\nonumber\\
\G^{r}_{v}=\frac{2\M_{v}}{r^{2}}=\frac{2m_{v}}{r^{2}}+\frac{2\phi_{v}\phi_{vv}}{r}~;~\nonumber\\
\G^{\t}_{\t}=\G^{\P}_{\P}=-\frac{\M_{rr}}{r}=-\frac{m_{rr}}{r}
\label{20}
\een

The `emergent' Einstein equation is
\ben
\G^{\mu}_{\nu} = \k {\cal T}^{\mu}_{\nu}
\label{21}
\een
where, $\k \equiv 8\pi G$ leads to the components ${\cal T}^{\mu}_{ \nu}$, which can be parametrized exactly as in ref.  \cite{husain,wang,maombi} in terms of the components $\g, \r$ and $\S$ given by
\ben
{\cal T}_{\mu\nu}={\cal T}_{\mu\nu}^{(n)} + {\cal T}_{\mu\nu}^{(m)}=
\left[\begin{array}{cccc}
(\g/2+\r) & \g/2 & 0 & 0	\\
\g/2 & (\g/2-\r) & 0 & 0	\\
0 & 0 & \S & 0	\\
0 & 0 & 0 & \S
\end{array}\right]\nonumber\\
\label{22}
\een
where ${\cal T}_{\mu\nu}^{(n)}=\g l_{\mu}l_{\nu}$;~ ${\cal T}_{\mu\nu}^{(m)}=(\r+\S)(l_{\mu}n_{\nu}+l_{\nu}n_{\mu})+\S\bar{G}_{\mu\nu}$
with $l_{\mu}$ and $n_{\mu}$ are two null vectors. Contractions of all indices are performed through the emergent metric $\bar{G}_{\mu \nu}$. The expressions for the three independent components are given by,
\ben
\g &=&\frac{2\M_{v}}{\k r^{2}}= \frac{2m_{v}}{\k r^{2}}+\frac{2\phi_v \phi_{vv}}{\k r} \label{23} \\
\r &=&\frac{2\M_{r}}{\k r^{2}}= \frac{2m_{r}}{\k r^{2}}+\frac{\phi_{v}^{2}}{\k r^{2}} \label{24} \\
\S &=& -\frac{\M_{rr}}{\k r}=-\frac{m_{rr}}{\k r}. \label{25}
\een

The energy conditions for 
the combination of Type-I and Type-II fluids energy momentum tensor ${\cal T}_{\mu\nu}$ as defined in \cite{haw-ellis} are as follow:

(a) The weak and strong energy conditions
\ben
\g\geq0~,~\rho\geq0~,~\S\geq0~~~(\g\neq0).
\label{26}
\een
(b) The dominant energy conditions
\ben
\g\geq0~,~\rho\geq\S\geq0  ~~~~~(\g\neq0).
\label{27}
\een

It is obvious that the above energy conditions (26) and (27) imposed on ${\cal T}_{\mu \nu}$ will, in turn, constrain $m(v,r)$ and $\phi(v)$ and their derivatives. Thus, 
\ben
\g > 0 &\Rightarrow & m_{v}+r\phi_v \phi_{vv} > 0 , \label{28} \\
\r > 0 & \Rightarrow & 2m_{r}+\phi_{v}^{2}>0, \label{29} \\
\S>0 & \Rightarrow & m_{rr} < 0 . \label{30}
\een

\section{Collapsing Scenario for the k-essence emergent Vaidya spacetime}
We note $K^{\mu}$ as the tangent to the nonspacelike geodesics with $K^{\mu}=\frac{dx^{\mu}}{d\l}$ where $\l$ is the affine parameter. The geodesic equation can be written as
\ben
K^{\nu}\nabla_{\nu}K^{\mu}=0~~and~~\bar{G}_{\mu\nu}K^{\mu}K^{\nu}=\b
\label{31}
\een
where $\b$ is a constant. $\b=0$ describes the null geodesics and $\b<0$ describes the timelike geodesics. The emergent geodesic equation (11) can be re-written in terms of $K^{\mu}$ as
\ben
\frac{dK^{\a}}{d\l}+\bar\Gamma ^{\a}_{\mu\nu}K^{\mu}K^{\nu}=0.
\label{32}
\een
Using the equations (32) and (15) the geodesic equations \cite{joshi3,maombi} have the forms:
\ben
\frac{dK^{v}}{d\l}+\Big[\frac{\M}{r^{2}}-\frac{\M_{r}}{r}\Big](K^{v})^{2}-r\Big[(K^{\t})^{2}+sin^{2}\t(K^{\P})^{2}\Big]&=&0\nonumber\\
\Rightarrow \frac{dK^{v}}{d\l}+\Big[\frac{\M}{r^{2}}-\frac{\M_{r}}{r}\Big](K^{v})^{2}-\frac{l^{2}}{r^{3}}&=&0\nonumber\\
\Rightarrow \frac{dK^{v}}{d\l}+\Big[\frac{m}{r^{2}}-\frac{m_{r}}{r}\Big](K^{v})^{2}-\frac{l^{2}}{r^{3}}&=&0,
\label{33}\nonumber\\
\een

\ben
\frac{dK^{r}}{d\l}+\frac{\M_{v}}{r}(K^{v})^{2}+\Big[\frac{\M}{r^{2}}-\frac{\M_{r}}{r}\Big]\Big[\Big(1-\frac{2\M}{r}\Big)(K^{v})^{2}-2K^{r}K^{v}\Big]\nonumber\\
-\Big(1-\frac{2\M}{r}\Big)r\Big[(K^{\t})^{2}+sin^{2}\t(K^{\P})^{2}\Big]=0\nonumber\\
\Rightarrow \frac{dK^{r}}{d\l}+\frac{\M_{v}}{r}(K^{v})^{2}-\frac{l^{2}}{r^{3}}\Big(1-\frac{2\M}{r}\Big)-\b\Big[\frac{\M}{r^{2}}-\frac{\M_{r}}{r}\Big]=0\nonumber\\
\Rightarrow \frac{dK^{r}}{d\l}+\Big(\frac{m_{v}}{r}+\phi_{v}\phi_{vv}\Big)(K^{v})^{2}-\frac{l^{2}}{r^{3}}\Big(1-\frac{2m}{r}-\phi_{v}^{2}\Big)\nonumber\\-\b\Big[\frac{m}{r^{2}}-\frac{m_{r}}{r}\Big]=0,~~~~~~~
\label{34}
\een

\ben
\frac{dK^{\t}}{d\l}+\frac{2}{r}K^{\t}K^{r}-sin\t~cos\t(K^{\P})^{2}&=&0,
\label{35}
\een

\ben
\frac{d}{d\l}(r^{2}sin^{2}\t K^{\P})&=&0.
\label{36}
\een
Following \cite{joshi3}, we use the relation 
\ben
K^{\P}=\frac{lcos\d}{r^{2}sin^{2}\t}~~,~~K^{\t}=\frac{l}{r^{2}}sin\d cos\P
\label{37}
\een
where $l$ and $\d$ are constant of integration. Here $l$ is the impact parameter and $\d$ is
the isotropy parameter such that $sin\P~tan\d=cot\t$.

We define $K^{v}$ \cite{joshi3,maombi} as
\ben
K^{v}=\frac{P(v,r)}{r}
\label{38}
\een
where $P(v,r)$ is an arbitrary function, then from the relation $\bar{G}_{\mu\nu}K^{\mu}K^{\nu}=\b$ gives
\ben
K^{r}=\frac{P}{2r}\Big(1-\frac{2m}{r}-\phi_{v}^{2}\Big)-\frac{l^{2}}{2Pr}+\frac{\b r}{2P}
\label{39}
\een

By differentiating the equation (38) with respect to $\l$ we get
\ben
\frac{dP}{d\l} &=& \frac{1}{r}\Big(r^{2}\frac{dK^{v}}{d\l}+P\frac{dr}{d\l}\Big)\nonumber\\
&=& \frac{P^{2}}{2r^{2}}\Big(1-\frac{4\M}{r}+2\M_{r}\Big)+\frac{l^{2}}{2r^{2}}+\frac{\b}{2}\nonumber\\
&=& \frac{P^{2}}{2r^{2}}\Big(1-\frac{4m}{r}+2m_{r}\Big)+\frac{l^{2}}{2r^{2}}+\frac{\b}{2}-\frac{P^{2}}{2r^{2}}\phi_{v}^{2}~~~~
\label{40}
\een

We check for the given  {\bf k-}essence emergent generalised Vaidya mass function (16), how the final fate of collapse is determined in terms of either a black hole or a naked singularity. If
there are families of future directed nonspacelike trajectories reaching faraway observers in spacetime and
terminating in the past at the singularity, we have a naked singularity forming as the collapse final state.
Otherwise an event horizon forms sufficiently early to cover the singularity in the absence of such families, we have a
black hole. 

Using equations (38) and (39), we have the equation for the radial null geodesics $(l=0,~\b=0)$ for the line element (15) is
\ben
\frac{dv}{dr}=\frac{2r}{r-2m(v,r)-r\phi_{v}^{2}}
\label{41}
\een
which has a singularity at $r=0,~v=0$ provided $\phi_{v}^{2}\neq 0$.

{\bf (a) Struture of the Central singularity for the emergent spacetime (15):}

In general, the above equation (41) can be written as \cite{maombi}
\ben
\frac{dv}{dr}=\frac{M(v,r)}{N(v,r)}
\label{42}
\een
with the singular point $r=0, v=0$, where both the above functions $M(v,r)$ and $N(v,r)$ vanish. 
In \cite{maombi}, they have proved, in general, the existence and uniqueness of the above form of differential equation (42) in the vicinity of this singularity. In this proof, they also have established the characteristic equation as
\ben
\chi^{2}-(A+D)\chi +AD-BC=0
\label{43}
\een
where $A=M_{v}(0,0)~,~B=M_{r}(0,0)~,~C=N_{v}(0,0)~,~D=N_{r}(0,0)$ and $AD-BC\neq 0$. The roots of the above equation (43) are
\ben
\chi=\frac{1}{2}\Big[(A+D)\pm \sqrt{(A-D)^{2}+4BC}\Big].
\label{44}
\een
The singularity is classified as a {\it node} if $(A-D)^{2}+4BC\geq 0$ and $BC>0$. Otherwise, it may be a center or focus.

Compairing equations (41) and (42), we have
\ben
M(v,r) &=& 2r,\nonumber\\N(v,r) &=& r-2\M(v,r) \nonumber\\&=& r-2m(v,r)-r\phi_{v}^{2}.
\label{45}
\een
At the central singularity ($v=0, r=0$), we define
\ben
\M_{0} &=& \lim\limits_{v\to 0, r \to 0}\M(v,r)~;~
\M_{v0} = \lim\limits_{v\to 0, r \to 0}\frac{\p}{\p v}\M(v,r)~;~\nonumber\\
\M_{r0} &=& \lim\limits_{v\to 0, r \to 0}\frac{\p}{\p r}\M(v,r)~;~
\phi_{v0} = \lim\limits_{v\to 0, r \to 0}\frac{\p}{\p v}\phi(v)~;~\nonumber\\
m_{v0} &=& \lim\limits_{v\to 0, r \to 0}\frac{\p}{\p v}m(v,r)~;~
m_{r0} = \lim\limits_{v\to 0, r \to 0}\frac{\p}{\p r}m(v,r).\nonumber\\
\label{46}
\een
Using equations (45) and (46), we get
$A=0,~B=2,~C=-2m_{v0}~ and ~ D=1-2m_{r0}-\phi_{v0}^{2}$. So, using equation (44), we have the roots of the characteristic equation (43) becomes
\ben
\chi=\frac{1}{2}\Big[(1-2m_{r0}-\phi_{v0}^{2})\pm \sqrt{(1-2m_{r0}-\phi_{v0}^{2})^{2}-16m_{v0}}\Big].\nonumber\\
\label{47}
\een

For the singular point $(r=0, v=0)$ to be a node, then the condition is
\ben
\Big(1-2m_{r0}-\phi_{v0}^{2}\Big)^{2}\geq16m_{v0},~
m_{v0}>0~and~\phi_{v0}^{2}>0.\nonumber\\
\label{48}
\een
Therefore, if the {\bf k-}essence emergent Vaidya mass function $\M(v,r)$ is chosen in such a manner that the condition (48) is satisfied, the singularity at the origin $(v=0, r=0)$ will be a node and outgoing nonspacelike geodesics can come out of the singlarity with a definite value of the tangent. 

Using equations (46) and (41), the linearized null geodesic equation near the central singularity can be written as
\ben
\frac{dv}{dr}=\frac{2r}{(1-2m_{r0}-\phi_{v0}^{2})r-2m_{v0}v}
\label{49}
\een
which is also singular at $v=0, r=0$.

{\bf (b) Existence of outgoing nonspacelike geodesics for the emergent spacetime (15):} 

We consider that the emergent mass function $\M(v,r)$ obeys the all physical energy conditions (26) and (27) with the constraints (28), (29) and (30) and exists the partial derivatives of the mass function which are continuous in the entire {\bf k-}essence emergent Vaidya spacetime and also obeys the conditions (48) at the central singularity. We find the condition for the existence of outgoing radial nonspacelike geodesics of this spacetime from the nodal singularity. To this aim, we consider $X$ as the tangent of these curves at the singularity. Following \cite{maombi,patil}, we determine the nature (a black hole or a naked singularity) of the collapsing solutions, for this, we also define
\ben
X_{0}=\lim\limits_{v\to 0, r \to 0}X=\lim\limits_{v\to 0, r \to 0}\frac{v}{r}
\label{50}
\een
Using equation (49), (50) and L'Hospital's rule, we get
\ben
X_{0}=\lim\limits_{v\to 0, r \to 0}\frac{v}{r}=\frac{dv}{dr}=\frac{2}{(1-2m_{r0}-\phi_{v0}^{2})-2m_{v0}(\frac{v}{r})}\nonumber\\
\label{51}
\een
which can be written as
\ben
X_{0}=\frac{2}{(1-2m_{r0}-\phi_{v0}^{2})-2m_{v0}X_{0}}.
\label{52}
\een
From (52) we get the values of $X_{0}$ as
\ben
X_{0}&=&\frac{(1-2m_{r0}-\phi_{v0}^{2})\pm \sqrt{(1-2m_{r0}-\phi_{v0}^{2})^{2}-16m_{v0}}}{4m_{v0}}\nonumber\\
=b_{\pm}
\label{53}
\een
where $b_{+}$ is the positive root and $b_{-}$ is the negative root of $X_{0}$.
For one or more real positive roots of $X_{0}$, the singularity may be locally naked if the null geodesic lie outside the trapped region. When there is no positive real root
of equation (53), the singularity is not naked because in that case there is no outgoing future
directed null geodesics from the singularity. Hence, in the absence of positive real roots,
the collapse ultimately leads to a black hole.

{\bf (c) Apparent horizon for the emergent spacetime (15):} 

During the evolution of collapse, the occurrence of a naked singularity or a black hole is usually decided by causal behaviour of the trapped surfaces. The boundary of the trapped surface region of the emergent spacetime is known as the apparent horizon. The apparent horizon for the {\bf k-}essence emergent Vaidya spacetime (15) is 
\ben
\frac{2\M(v,r)}{r}=1
\nonumber\\
\Rightarrow \frac{2m(v,r)}{r}+\phi_{v}^{2}=1.
\label{54}
\een
On following \cite{maombi}, we calculate the slope of the emergent apparent horizon in the following manner:
\ben
2\Big(\frac{\p\M}{\p v}\Big)\Big(\frac{dv}{dr}\Big)_{AH}+\frac{2\p\M}{\p r}=1.
\label{55}
\een
From the above equation (55), we obtain the slope of the emergent apparent horizon at the central singularity $(v\rightarrow0,~r\rightarrow0)$ as
\ben
\Big(\frac{dv}{dr}\Big)_{AH}=\frac{1-2m_{r0}-\phi_{v0}^{2}}{2m_{v0}}
\label{56}
\een

Thus the sufficient conditions for the existence of a locally naked singularity for the collapsing {\bf k-}essence emergent Vaidya spacetime is attained.
The emergent mass function $\M(v,r)$ obeys all the energy conditions and constraints (28,29,30) and is differentiable in the entire emergent spacetime (15), if the following conditions are satisfied:
(a) The limits of the partial derivatives of the mass function $\M(v,r)$ at the central singularity obey the conditions (48).
(b) One or more positive real roots $X_{0}$ of equation (53) must occur.
(c) At least one of the positive real roots is less than $(\frac{dv}{dr})_{AH}$ at the central singularity. 

In\cite{maombi}, they have shown that the values of  $X_{0}=b_{\pm}$ must lie below the value of $(\frac{dv}{dr})_{AH}$ and there exist an open set of parameter values for which the singularity is locally naked for the case of generalized Vaidya spacetime.

{\it For the case of {\bf k-}essence emergent Vaidya spacetime, we can also show that the values of $X_{0}$ are always below the value of $(\frac{dv}{dr})_{AH}$ provided $m_{v0}>0$. This statement can be proved as:
\ben
X_{0} < (\frac{dv}{dr})_{AH}~;~
\Rightarrow \frac{y\pm\sqrt{y^{2}-16m_{v0}}}{4m_{v0}} &<& \frac{y}{2m_{v0}}\nonumber\\
\Rightarrow y^{2}-16m_{v0}<y^{2}~;~
\Rightarrow m_{v0}>0
\een
where $y=1-2m_{r0}-\phi_{v0}^{2}$.
Therefore, we can also infer that an open set of parameter values must exist for which the singularity is locally naked for the case of  {\bf k-}essence emergent Vaidya spacetime.}

As an example, we consider the value of the kinetic energy of the {\bf k-}essence scalar field at the central singularity is $\phi_{v0}^{2}=0.7$ and $m_{v0}=0.0015$. Figure 1 ($-2.68\leq m_{r0}\leq -2.64$) and figure 2 ($-3\leq m_{r0}\leq 0.01$) show that the values of $X_{0}=b_{\pm}$ become smaller than those appears of $(\frac{dv}{dr})_{AH}$ and thus for our case, the above statement is true as \cite{maombi}.

\begin{figure}[h]
 \includegraphics[width=7cm, height=6cm]{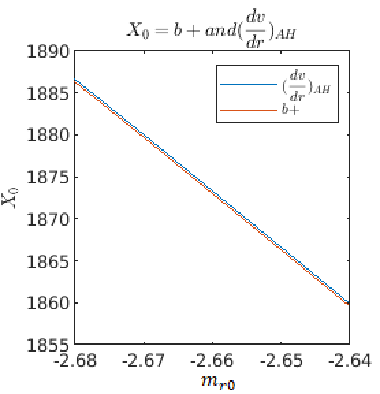}
 \caption{(Color online) Variation of $X_0=b_{+}$ (red) and $\left(\frac{dv}{dr}\right)_{AH}$ (blue) with $m_{r0}$ at fixed value $m_{v0}= 0.0015$}\label{variation1}
\end{figure}

\begin{figure}[h]
 \includegraphics[width=7cm, height=6cm]{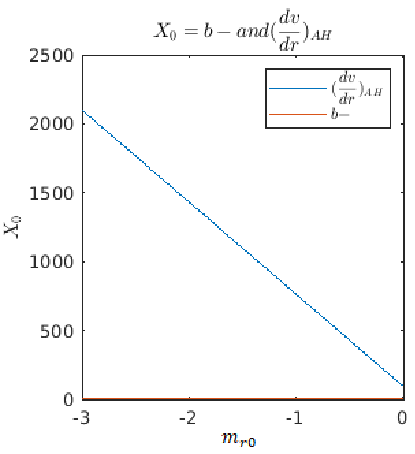}
 \caption{(Color online) Variation of $X_0=b_{-}$ (red) and $\left(\frac{dv}{dr}\right)_{AH}$ (blue) with $m_{r0}$ at fixed value $m_{v0}= 0.0015$}\label{variation2}
\end{figure}

{\bf (d) Strength of singularity for the emergent spacetime (15):}

By following \cite{maombi,patil,tipler,ghosh,clarke}, we can highlight that a singularity $(r=v=\l=0)$ would be strong if the condition 
\ben
\lim\limits_{\l\to 0}\l^{2}\psi=\lim\limits_{\l\to 0}\l^{2}\bar{R}_{\mu\nu}K^{\mu}K^{\nu}>0
\label{58}
\een
is satisfied, where $\bar{R}_{\mu\nu}$ is Ricci tensor and we define $\psi=\bar{R}_{\mu\nu}K^{\mu}K^{\nu}$ as a scalar of the {\bf k-}essence emergent Vaidya spacetime. Mention that this scalar $\psi$ is not {\bf k-}essence scalar field also not coupled with the background gravitational metric $g_{\mu\nu}$. Using equations (18), (38) and (39), we get
\ben
\l^{2}\psi=\Big(2\M_{v0}\Big)\Big(\frac{P\l}{r^{2}}\Big)^{2}.
\label{59}
\een
If we assume that $P(v,r)\neq 0,\infty$ and using L'Hospital rule, we get
\ben
\lim\limits_{\l\to 0}\l^{2}\psi=\Big(2\M_{v0}\Big)\lim\limits_{\l\to 0}\Big(\frac{P\l}{r^{2}}\Big)^{2}\nonumber\\
=\Big(2\M_{v0}\Big)\lim\limits_{\l\to 0}\Big(\frac{Pd\l}{2rdr}\Big)^{2}.
\label{60}
\een

Using equation (38), $\frac{P}{r}=\frac{dv}{d\l}$, we get

\ben
\lim\limits_{\l\to 0}\l^{2}\psi
=\Big(2\M_{v0}\Big)\Big(\frac{1}{2}\frac{dv}{dr}\Big)^{2}=\Big(2\M_{v0}\Big)\frac{1}{4}X_{0}^{2}\nonumber\\
=\Big(2m_{v0}\Big)\frac{1}{4}X_{0}^{2}
\label{61}
\een
where at the limit $(v\rightarrow0,r\rightarrow0)$, $\M_{v0}=m_{v0}$ irrespective of the {\bf k-}essence scalar field. But from equation (53) $X_{0}$ is found to be dependent on limiting value of the {\bf k-}essence scalar field. {\it Therefore,  
with the conditions (48) and suitable choice of the emergent mass function (16), it can be shown that
\ben
\lim\limits_{\l\to 0}\l^{2}\psi=\Big(2\M_{v0}\Big)\frac{1}{4}X_{0}^{2}>0.
\label{62}
\een
If this condition is satisfied for some real and positive roots of $X_{0}$, the naked singularity of the {\bf k-}essence emergent Vaidya spacetime is strong. Otherwise i.e., if there is no positive real root of $X_{o}$, we can conclude that there is no outgoing future
directed null geodesics from the singularity comes into the picture i.e., the collapse ends into a black hole.}

\section{Background gravitational metric to be some special subclasses of generalized Vaidya metric}

In this section, we evaluate the values of the three independent components ($\g, \rho, \S$) of energy-momentum tensor $\cal T_{\mu\nu}$ using equations (23,24,25) and the equations of tangents (52) to the null geodesics at the central singularity $(v\rightarrow0,r\rightarrow0)$ for some special subclasses of the generalized Vaidya backgroud spacetime with the specific mass function $\M(v,r)$.

A physical situation where a radial influx of
null fluid in an initially empty region of the Minkowski spacetime \cite{maombi,dadhich} is taken into consideration. The first shell arrives at $r = 0$ at time $v = 0$ and the final at $v = T$. A central singularity of
growing mass is developed at $r = 0$. For $v < 0$ we have $m(v) = q(v) = 0$, i.e., an empty Minkowski metric appears, and for $v > T$, $m_{v}(v) = q_{v}(v) = 0$, $m(v)$ and $q^{2}(v)$ are
positive definite. The metric for $v = 0$ to $v = T$ is the
{\bf k-}essence emergent Vaidya spacetime, and for $v > T$ we have $m(v)=M_{0},~q^{2}(v)=q_{0}^{2}$ where $M_{0}$ and $q_{0}^{2}$ are some positive constants. In the limit $v>T$ the emergent gravity metric is analogues to the Barriola-Vilenkin type spacetime for the Schwarzschild background \cite{gm1} and Robinson-Trautman type spacetime for the Reissner-Nordstrom background \cite{gm2} where $\phi_{v}^{2}=constant$ and is also analogues to {\bf k-}essence generalization of Vaidya spacetime \cite{gm4} where $\phi_{v}^{2}$ is function of $v$ only.

{\bf a) The usual Vaidya background:}
In this case, we consider the mass function \cite{papa,maombi} 
\ben
m(v,r)=m(v).
\label{63}
\een
The {\bf k-}essence mass function (16) becomes
$\M(v,r)=m(v)+\frac{r}{2}\phi_{v}^{2}$ and defines the mass function $m(v)$ as

\ben
m(v)=\begin{cases}
 0 &if~v < 0, \\
\frac{1}{2}\e v
& if~0\leq v \leq T,\\
M_0 & if~v > T.
\end{cases}
\label{64}
\een
where $\e$ is an arbitrary constant.
For this case, the values of $\g,~\rho~and~\S$ are

\ben
\g=\frac{2m_{v}}{\k r^{2}}+\frac{2\phi_{v}\phi_{vv}}{\k r};~
\rho=\frac{\phi_{v}^{2}}{\k r^{2}};~ \S=0.
\label{65}
\een
For the above values of $\g,~\rho~and~\S$, the energy-momentum tensor $\cal T_{\mu\nu}$ belongs to type-II
class \cite{haw-ellis} which has a double null vector. 
The tangent equations (53) to the null geodesics at the central singularity $(v\rightarrow0,r\rightarrow0)$ for the range $0\leq v \leq T$ are
\ben
X_{0}=\frac{(1-\phi_{v0}^{2})\pm \sqrt{(1-\phi_{v0}^{2})-8\e}}{2\e}.
\label{66}
\een
If we consider $\phi_{v0}^{2}=0.7$, the above equation gives the positive values of $X_{0}$ for all values of $\e$ in the range $0<\e \leq \frac{3}{80}$. Also, from (62) we can show that
$\lim\limits_{\l\to 0}\l^{2}\psi=\frac{1}{4}\e X_{0}^{2}>0$ for all values of $X_{0}$ in the above range of $\e$. Therefore, we can say that the singularity is strong.

{\bf b) The charged Vaidya background}:
On the basia of\cite{maombi,wang,dadhich}, we conider the mass function
\ben
m(v,r)=f(v)-\frac{q^{2}(v)}{2r}
\label{67}
\een
where $f(v)$ and $q(v)$ are arbitrary functions of mass and electric charge respectively. Thus the {\bf k-}essence mass function is $\M(v,r)=f(v)-\frac{q^{2}(v)}{2r}+\frac{r}{2}\phi_{v}^{2}$.
So the values of $\g,~\rho~and~\S$ are
\ben
\g=\frac{2f_{v}}{\k r^{2}}-\frac{2qq_{v}}{\k r^{3}}+\frac{2\phi_{v}\phi{vv}}{\k r}~;\nonumber\\
\rho=\frac{q^{2}}{\k r^{4}}+\frac{\phi_{v}^{2}}{\k r^{2}}~;~
\S=\frac{q^{2}}{\k r^{4}}.
\label{68}
\een
which have satisfied the weak and strong energy conditions (26). Now we define $f(v)$ and $q(v)$ as \cite{maombi,dadhich}
\ben
f(v)=\begin{cases}
 0 &if~v < 0, \\
\e v(\e>0) & if~0\leq v \leq T,\\
f_{0}(>0) & if~v > T.
\end{cases}
\label{69}
\een
and
\ben
q^{2}(v)=\begin{cases}
 0 &if~v < 0, \\
\mu^{2} v^{2}(\mu^{2}>0) & if~0\leq v \leq T,\\
q_{0}^{2}(>0) & if~v > T.
\end{cases}
\label{70}
\een
then the tangent equation (52) in the range $0\leq v \leq T$ becomes
\ben
\mu^{2}X_{0}^{3}-2\e X_{0}^{2}+(1-\phi_{v0}^{2})X_{0}-2=0.
\label{71}
\een

If the singularity is naked, equation (71) must have one or more positive roots of $X_{0}$ considering $\phi_{v0}^{2}=0.7$,
i.e. at least one outgoing geodesic that will terminate in the past at the singularity.
Hence in the absence of positive roots, the collapse will always lead to a black hole.
Thus, the occurrence of positive roots implies that the strong CCC is violated, though not necessarily being the weak one. It has been shown that the central shell
focusing singularity is at least locally naked if this algebraic equation (71) admits one or more
positive roots \cite{joshi2}. When there is no positive root to (71), the central singularity is
not naked because in that case there is no outgoing future directed null geodesics emerging from the singularity. Hence in the absence of positive roots, the collapse
will always lead to a black hole. As an example, we consider $\mu^{2}=0.001, \e=0.01~and~ \phi_{v0}^{2}=0.7$ then one of the roots of the equation (71) is $10$ which is positive. Thus the condition for a strong singularity is satisfied.

{\bf c) The charged Vaidya  deSitter background}:
Here we chose the mass function as \cite{bees}
\ben
m(v,r)=f(v)-\frac{q^{2}(v)}{2r}+\frac{\L r^{3}}{6}
\label{72}
\een
where $\L\neq 0$ is the cosmological constant and the {\bf k-}essence emergent Vaidya mass function is $\M(v,r)=f(v)-\frac{q^{2}(v)}{2r}+\frac{\L r^{3}}{6}+\frac{r}{2}\phi_{v}^{2}$.
Thus the values of $\g,~\rho~and~\S$ are
\ben
\g=\frac{2f_{v}}{\k r^{2}}-\frac{2qq_{v}}{\k r^{3}}+\frac{2\phi_{v}\phi{vv}}{\k r}~;\nonumber\\
\rho=\frac{q^{2}}{\k r^{4}}+\frac{\phi_{v}^{2}}{\k r^{2}}+\frac{\L}{\k}~;~
\S=\frac{q^{2}}{\k r^{4}}-\frac{\L}{\k}
\label{68}
\een
which have follow the energy condition (26). We observe that the tangent equation (52) for this case is similar to that of charged Vaidya backgroud case (71). So we can say that the singularity is also strong.

\section{Conclusion}
Based on the Dirac-Born-Infeld Lagrangian, the gravitational collapse of null fluids is analyzed in the the {\bf k-}essence emergent Vaidya spacetime. The analysis is presented in the context of cosmic censorship hypothesis with the {\bf k-}essence emergent Vaidya mass function. We develop the Einstein tensor and the energy conditions for the combination of Type-I and Type-II fluids energy momentum tensor for the {\bf k-}essence emergent Vaidya spacetime.
We have studied the classification of non-spacelike geodesic for the {\bf k-}essence emergent Vaidya spacetime that connect the naked singularity in the past, which is a strong curvature singularity
in a stronger sense.
The central singularity being a node allows outgoing non-spacelike
geodesics to come out of the singlarity with a definite value of the tangent. The positive value of the tangent implies that the strong
CCC is violated i.e., the singularity is locally naked, though not necessarily being the weak one.
If the value of tangent is not positive, the central singularity is not naked because, in that case, there is no outgoing future directed
null geodesics emerging from the singularity i.e., the collapse will always
lead to a black hole.
We also have the same type of results as Mkenyeleye et. al. \cite{maombi} in the presence of {\bf k-}essence scalar field for the {\bf k-}essence emergent Vaidya spacetime for the collapsing scenario.

\vspace{0.1in}

{\bf Acknowledgement:}
I would like to thank Pradipta Panchadhyayee, Department of Physics and Bivash Majumder,  Department of Mathematics,
Prabhat Kumar College, Contai, Purba Medinipur, West Bengal, India for valuable suggestions and discussions. The author would like to thank the
referees for illuminating suggestions to improve the manuscript.

\end{document}